# Tracking electricity losses and their perceived causes using nighttime light and social media

Samuel W Kerber[1], Nicholas A Duncan[1], Guillaume F L'Her[1,2], Morgan Bazilian[3], Chris Elvidge[3], Mark R Deinert[1,2,3*]

[1]Department of Mechanical Engineering, The Colorado School of Mines, Golden, CO
[2]Nuclear Science and Engineering, The Colorado School of Mines
[3]Payne Institute for Public Policy, The Colorado School of Mines

**\*Corresponding Author**. Mark R. Deinert, Department of Mechanical Engineering, Payne Institute for Public Policy, The Colorado School of Mines. Email: mdeinert@mines.edu

**Abstract**. Urban environments are intricate systems where the breakdown of critical infrastructure can impact both the economic and social well-being of communities. Electricity systems hold particular significance, as they are essential for other infrastructure, and disruptions can trigger widespread consequences. Typically, assessing electricity availability requires ground-level data, a challenge in conflict zones and regions with limited access. This study shows how satellite imagery, social media, and information extraction can monitor blackouts and their perceived causes. Night-time light data (in March 2019 for Caracas, Venezuela) is used to indicate blackout regions. Twitter data is used to determine sentiment and topic trends, while statistical analysis and topic modeling delved into public perceptions regarding blackout causes. The findings show an inverse relationship between nighttime light intensity. Tweets mentioning the Venezuelan President displayed heightened negativity and a greater prevalence of blame-related terms, suggesting a perception of government accountability for the outages.

**Introduction.** Urban communities are complex systems with strongly coupled networks of critical infrastructure, regulation, governance, and social networks. Reliable electricity is central to all of these and its failure can have potentially catastrophic cascade effects[1]. While natural hazards often cause disruptions in energy systems[2], past work has also shown that their performance can also be affected by conflict events[3–8]. Whether these systems are functioning or not can also be used as an indicator of disruption by social unrest, the occurrence of hazard events in a region, or infrastructure frailty. Unfortunately, quantifying electricity availability typically requires ground level observations. This can be difficult to impossible in conflict zones and regions that have been hit by natural disasters or to which access is restricted. However, systems that allow for remote sensing of power systems hold promise for increasing the range of locations that can be monitored as well as reducing the costs associated with tracking outages[1].

Remote sensing for urban environments had its genesis in military applications. Aerial reconnaissance has been used in conflict monitoring since 1784 when balloons were used for scouting in the Battle of Fleurus[9], and by the 1870s France had a balloon corps for this purpose[10]. In World War I, balloons and airplanes were used for battlefield surveillance[11]. Long-range reconnaissance was employed in the cold war using high altitude airplanes [12]. Drones were proposed in 1915 by Nikola Tesla on the possibility of unmanned, armed aircraft, and in 1919, Elmer Sperry used such a drone to destroy a captured German ship. Common use occurred later and became prevalent with the "Pioneer" drone in 1985, which would be used in more than 300 combat missions for Operation Desert Storm in 1991[13]. Predator drones, an unmanned aerial vehicle, were first used for surveillance and strikes in Operation Enduring Freedom in 2001[14]. In 2008, drones recorded more than 135,000 flying hours over Afghanistan and Iraq, with plans to triple the military inventory of drones by 2020[13]. Unmanned drones are now routinely used for aerial surveillance[15].

Satellite tracking and observation began in 1957[16] with an effort to catalog satellites by the National Space Surveillance Control Center (NSSCC) with the first satellite used for long-range communication in 1958[17]. 1962 saw the first observation efforts to measure sea ice by satellite, and the US Navy's involvement carried implications for the national security agenda[18]. In 1968, communications and intelligence satellites were launched by the United States specifically for military surveillance needs. By the 1990s, satellites were equipped with Global Positioning Systems which allowed detailed location information on the acquired images[17]. Satellite imaging has found widespread use in conflict monitoring. Studies of the Syrian civil war have employed the use of satellites to track damage to infrastructure, damage to historical sites, track the movements of refugees, and have been used to show how blackouts have progressed through the nation[19,20]. They have been used by non-governmental organizations in places like Myanmar to track damaged and burned buildings after rioting[21].

Satellite imaging can help with tracking power systems[22] and the availability of electricity in particular[23]. The Visible Infrared Imaging Radiometer Suite (VIIRS) on the NOAA-20 weather satellite provides detailed images of the world at night, including lights radiating from cities. The VIIRS satellite orbits the earth twice per night, in a polar orbit, taking photos on each pass and detecting available night time light with each pass. The National Oceanic and Atmospheric Administration (NOAA) generates images from the data which are reported online as daily or monthly mosaics. Li and Huang (2015) used this data to track the insurgency in Northern Iraq in 2014 by watching lights go out or turn back on[24]. The same approach was used to track conflict during the Syrian civil war[25]. More recently, satellite imaging of nighttime light has found use in understanding the reliability of electricity systems[26]. Mirza et al (2020) used nighttime light to measure global inequality[27] and McCallum et al (2022) used it to measure global well-being[28]. Zhang et al (2020) used the VIIRS system to show that Colombia had a decrease in nighttime lights related to power outages between 2012 and 2018[29]. Schweikert et al (2022) used data on nighttime light and cell phone mobility to study their relationship to satellite derived measurements of air pollution during the early phases of the SARS-CoV-2 pandemic[30].

While satellite imaging can be used to tell where and when lights have gone on or off, it cannot say why. For this, social media can sometimes provide time stamped information on an event or topic over a wide spread population. Twitter was used by terrorists for situational awareness in the Mumbai Terror Attacks in 2008[31]. It has also been used for crime prediction[32,33] as well as for monitoring of natural disasters and other events[34]. Live posts and historical archives of social media sites allow researchers and reporters to corroborate reports generated from government agencies and provide photographic evidence of conflicts without the need for onsite photographers from newspapers or watchdog groups. Tufekci (2014) discusses the importance of the rise of *latent data*, by which "campaigns can now capture actual utterances of people as they talk about a wide variety of topics of interest to them"[35]. The latent data stored in social databases can express how populations are reacting to and conversing about responses to the event, the event itself, and recovery from the event at the ground level. A growing body of studies have used such data to assess public sentiment toward various topics. These data have been used to monitor ongoing or past events, such as hurricanes[36,37], earthquakes[38], floods[39,40], epidemics[41,42], and more. Importantly, sentiment analysis using natural language processing allows for standardized quantification of emotional states from text[43]. This approach has been validated against conventional methods such as the Gallup–Sharecare Well-Being Index survey, which is considered a standard for measurement of subjective well-being[44,45].

Multiple studies have used satellite imaging and social media to monitor conflicts and natural hazards. Several recent studies have begun to explore the fusion of these data streams to fill in geospatial gaps or corroborate satellite observations, e.g., Hultquist[46] and Li[47]. Currently missing from this body of work is the use of information extraction algorithms with social media data to understand the underlying causes for events or failures in response. Here, we show that nighttime light can be used in combination with twitter data and information extraction to not only monitor power systems but also understand public reaction to outages. After blackouts in Venezuela in early March, 2019, massive protests gripped the nation. We analyze the correlation between nighttime light in Caracas Venezuela, during March that year, and blackout related tweets for this region. Keywords were used to attribute the cause of the blackouts. The approach provides a new method to monitor power systems in urban environments and understand factors that affect their public perception. Remote sensing combined with machine learning can help to create fully automated processes where monitoring infrastructure disturbances can be identified along with public response, sentiment, and blame.

**Results and discussion.** Nighttime light and Twitter data for all of March 2019 are shown in Fig. 1. The Spearman's rank correlation coefficient for these data is -0.375, p-value = 0.049, which shows that nighttime light and blackout related tweets are anticorrelated. This makes intuitive sense as people would not be expected to post to social media during a power outage but when power was regionally available again.

The data acquired via the Twitter API revealed many facets of the public response to the blackouts. Tweets from the 8$^{th}$ and 10$^{th}$, the biggest of the March blackouts, show that Caracas blamed the Venezuelan government for the outages, called for protests, asked the outside world for help to expose the problem, and for water and basic human needs. The API data revealed how widespread the public discourse had become, and showed staggering rates of conversation surrounding the topic, as well as how the topic evolved over time. A total of 2326 tweets were isolated for Venezuela of which 752 were specific to Caracas. The highest tweet per day rate was calculated to be near 80,000 on the 8th, and only slightly lower at 60,000 on the 10th.

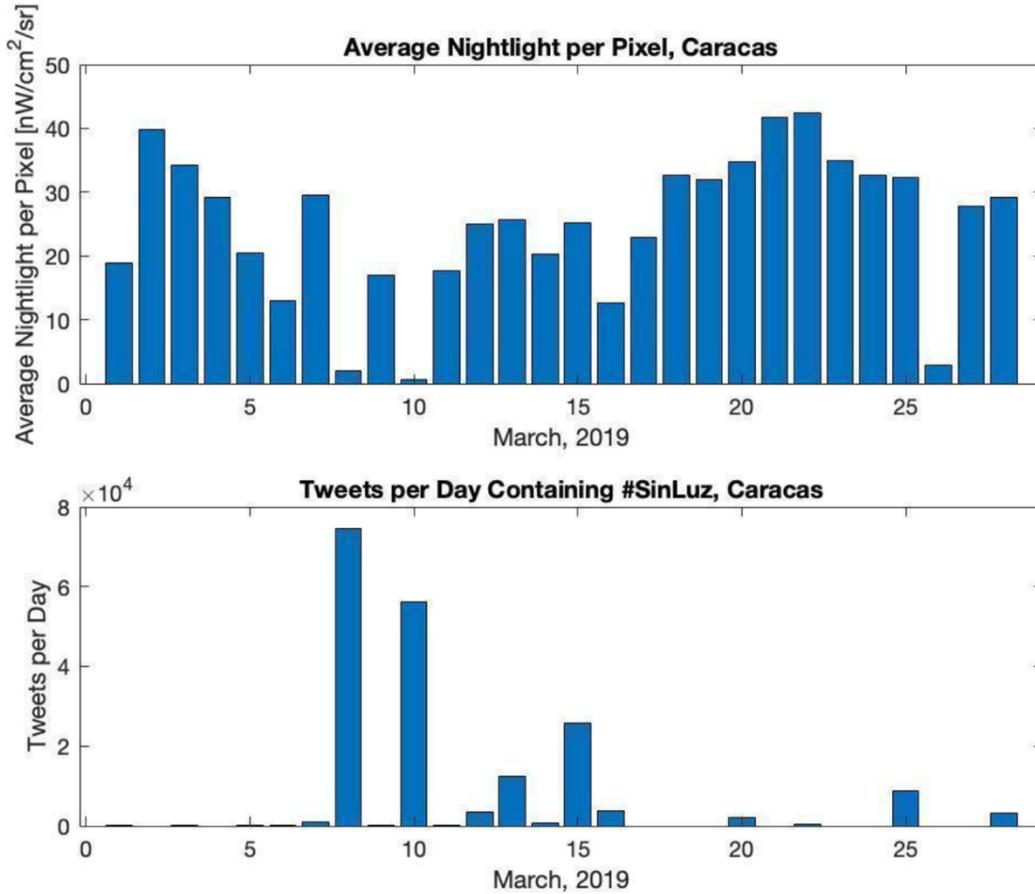

**Figure 1. Cross correlation between night lights and tweet rates within Caracas**. There is strong negative correlation with 0 lag which shows that #sinluz tweets were highest during peak low light times.

A word cloud was generated from the tweets. Superfluous words, such as "and," "an," the," "on," etc. were not included. The remaining words were translated to English only after they were counted to limit any errors due to translation. Word clouds were generated, with the size of the word correlated to the number of counts. A word cloud was generated for all tweets that do not contain the word "Maduro," Fig. 2 (left). Tweets over the time period were then isolated to include only the posts that contain the word "Maduro" in order to gain further insight on how the population was responding to the government's handling of the outages. As in the previous word cloud, superfluous words were removed, and the occurrence of individual words was used to generate the figure. The words were then translated, and a word cloud was generated, with the size of words corresponding to the number of occurrences of that word. The Maduro-specific word cloud can be found in Fig. 2 (right).

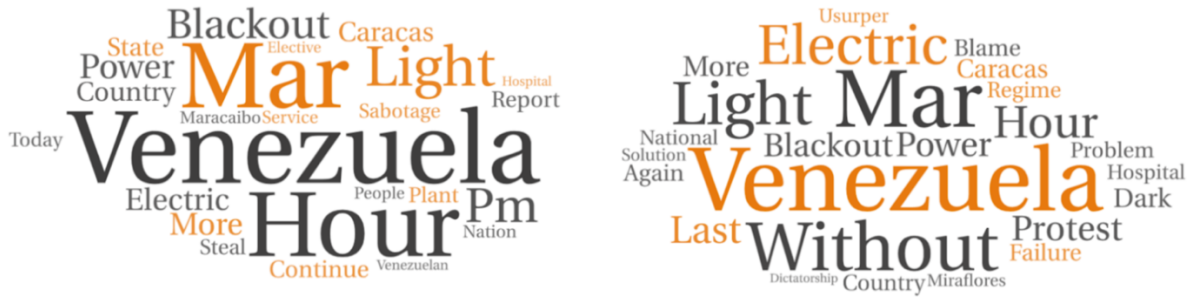

**Figure 2. Word Cloud of Tweets.** (left) shows a word cloud generated from all tweets that do not contain the word "Maduro". (Right) is a word cloud generated from tweets specifically containing the word "Maduro" over the same time frame. Each cloud contains 25 of some of the most used words, excluding common words and specific words outlined in Supplemental Information, Note S1, Excluded Word Cloud Words.

While there is considerable overlap between the two word clouds in Fig. 2, some differences are clear. In both, the commonsense words appear such as "electric," "power," "light" etc. However, on the left side we see words like "steal," "service," "report," "plant," "people" indicating conditions on the ground. This is compared to the tweets that do contain Maduro, in which words like "blame," "regime," "usurper," "failure," and "dictatorship" appear. The frequency of "blame," and synonyms for it, is also significantly different between tweets with and without "Maduro." Here, instances of "Blame," "Irresponsible," "Responsible," Responsibility," Irresponsibility," accountable," fault," guilty," and "condemning" were tallied for the tweets geolocated to Venezuela. For tweets containing "Maduro" 14.45% had at least one of the above blame words. For tweets that did not contain "Maduro" only 2.22% contained one or more of the above blame words, a statistically significant difference with a binomial test ($p < 0.001$). See Supplementary Information, Note 2, Statistics.

Tweets with and without 'Maduro' were isolated for Venezuela as a whole and for Caracas in particular. Sentiment analysis was performed to determine how people in these regions were responding to the blackouts with 0 indicating the most negative sentiment and 1 the most positive. Table 1 summarizes the results. In all cases, as expected, sentiment is extremely negative. The data shows that a higher proportion of tweets containing 'Maduro' are in negative categories than is the base for ones excluding 'Maduro.' However, enough tweets to determine the statistical significance of these differences only exist in the $0 - < 0.1$ bin. Two non-parametric tests were used, the two-tailed Mann-Whitney U test and the Wilcox Signed Rank test. For Venezuela as a whole, the Mann-Whitney U test showed tweets with and without 'Maduro' to be statistically different with a p-value of 0.004 and Wilcox Signed Rank test gave a p-value = 0.006. For tweets coming from Caracas, the lowest sentiment bin again has a higher proportion of tweets when they contain 'Maduro.' Both the Mann-Whitney U and Wilcox Signed Rank tests showed tweets with and without 'Maduro' to be statistically different with a p-value of 0.002. These results suggest that at least some degree of blame for the outages is associated with Maduro and/or his government.

| Sentiment Bin | Venezuela Tweets with Maduro | | Caracas Tweets with Maduro | | Venezuela Tweets without Maduro | | Caracas Tweets without Maduro | |
|---|---|---|---|---|---|---|---|---|
| | # Tweets | Percent of tweets | # Tweets | Percent of tweets | # Tweets | Percent of tweets | # Tweets | Percent of tweets |
| 0 - < 0.1 | 186 | 76.9 | 52 | 75.4 | 1474 | 70.7 | 492 | 72.0 |
| 0.1 - < 0.2 | 7 | 2.9 | 2 | 2.9 | 114 | 5.5 | 41 | 6.0 |
| 0.2 - < 0.3 | 12 | 5.0 | 1 | 1.4 | 102 | 4.9 | 33 | 4.8 |
| 0.3 - < 0.4 | 5 | 2.1 | 1 | 1.4 | 36 | 1.7 | 11 | 1.6 |
| 0.4 - < 0.5 | 10 | 4.1 | 5 | 7.2 | 137 | 6.6 | 39 | 5.7 |
| 0.5 - < 0.6 | 4 | 1.7 | 1 | 1.4 | 22 | 1.1 | 10 | 1.5 |
| 0.6 - < 0.7 | 0 | 0.0 | 0 | 0.0 | 31 | 1.5 | 8 | 1.2 |
| 0.7 - < 0.8 | 1 | 0.4 | 0 | 0.0 | 40 | 1.9 | 12 | 1.8 |
| 0.8 - < 0.9 | 11 | 4.5 | 6 | 8.7 | 49 | 2.4 | 15 | 2.2 |
| 0.9 - < 1.0 | 6 | 2.5 | 1 | 1.4 | 79 | 3.8 | 22 | 3.2 |

**Table 1. Percentage of tweets with sentiment score in ranges from 0 to 1.** All tweets excluding "Maduro" for Venezuela and those from Caracas specifically are shown as well as ones that contain the word "Maduro" for these regions. All Tweets include #SinLuz.

The results show a clear disdain, particularly at Maduro, during this period of time. This has important policy implications. The power of social media allows a policy maker to immediately, in real time, monitor the sentiment of a population during ongoing events. If specific keywords can be found to describe the event, policy makers may be able to tailor critical responses to the needs most relayed via social posts, such as needs in areas for food, water, or medical aid. Coupling this with remote sensing capabilities can narrow the scope of the social analysis to only affected areas, thus reducing noise from areas beyond the impact zone. During large scale events, this is useful to determine what the people in the thick of it need most. Furthermore, the sentiment analysis and investigations into these posts allows public facing entities to tailor responses, aid, and future solutions more effectively. It may also allow governments insight into how to better handle these events. From a purely political standpoint, this analysis may also be used to judge approval of actions, perceived blames, and effectiveness of messaging.

Natural language processing could also be used to mine social media posts for information about what other networks were disrupted by power outages. This would provide a remote sensing method to understand the interdependency of other networks (e.g., transportation, health care, governance, etc.) on electricity supply. This type of information is critical to the development of models that can predict the downstream effects of electricity loss. However, additional methods to geolocate tweets (beyond the user account giving a location) might be needed to ensure that results are statistically meaningful.

There are several limitations with the current work. The location data that was used to isolate tweets originating from Venezuela and Caracas was input by the Twitter user on their account. Since only a few percent of all Twitter accounts give a corresponding location for the user, this presents a limitation in geolocation. In particular, more tweets from Caracas would bolster the sentiment analysis which here was measured using a Spanish specific natural language processing algorithm. Tweets also do not necessarily represent a true cross-sectional view of sentiment within blackout regions because only people engaged over social media are captured by the data. However, validation studies suggest that even with this limitation, social media are representative for quantification of emotional states and subjective well-being (41)(42)(43). Better sentiment results might be obtained by using Sentiment Spanish as a pre-trained algorithm with additional specific training for sentiment analysis power outages and the perception of the Venezuelan government. While both Mann-Whitney U test and the Wilcox Signed Rank tests showed a statistically significant drop in sentiment for tweets containing 'Maduro,' correlation is not causation. However, these results are also consistent with the increase in blame word frequency seen in tweets containing "Maduro" relative to those that do not, and this too was found to be statistically significant. Even so, a larger geolocated tweet volume could allow machine learning approaches where training is done specifically to establish causality without the need to rely on sentiment and blame words as proxies. With the advent of large language models, approaches such as this are likely on the horizon. Unfortunately, Twitter discontinued its academic API in June of 2023 and dramatically limited the volume of tweets that could be accessed at a reasonable cost. Extensions to this, and other research, will require alternative sources of large-scale social media data unless Twitter reinstates an academic research access.

**Conclusions**. Past work has shown that satellite imaging can be used to monitor nighttime light and determine where outages occur. Past work with social media data has shown that it can be used to increase the spatial resolution for blackout regions. Here we show that these two data streams can also be used in combination to help understand both where blackouts are occurring and whom the public might assign blame to. Satellite images clearly showed nights where no significant light emanated from Caracas during the spring of 2019. Sentiment analysis of tweets showed them to be more negative when they mention "Maduro" then when they do not (p=0.002). Word frequency in tweets isolated to the Caracas blackouts was found to shed insight into public reaction and perception. The words 'Responsible' and 'unforgivable' were among the most used words in tweets containing "Maduro," with the former being the 10th most common word and the latter being the 15th most common word in that set of tweets. The frequency of "blame," and synonyms for it, was also considerably higher in tweets that contained "Maduro" relative to those that did not (p < 0.001). All of these data indicate specific disapproval of Maduro in the context of the blackouts and their aftermath and as a source of blame. Word frequency was also found to be an indicator of perceived instances of conflict (looting, chaos, sabotage, repression were also among the most common words in blackout related tweets).

**Methods.** VIIRS nighttime images were used to generate an average pixel value for the Caracas region for each night during March 2019. The twitter API was used to pull tweets containing the keyword #sinluz for each day during this period. These were geospatially filtered to isolate only tweets from Venezuela and Caracas by using the location data the account holder set. The tweets were tallied by day. Tweets were processed to give each a sentiment score. These scores were tallied and analyzed to determine sentiment distributions. Tweets were additionally filtered for references to the President of Venezuela and sentiment distributions with and without this reference were also scored to determine the sentiment distributions. Figure 3 summarizes the methodology.

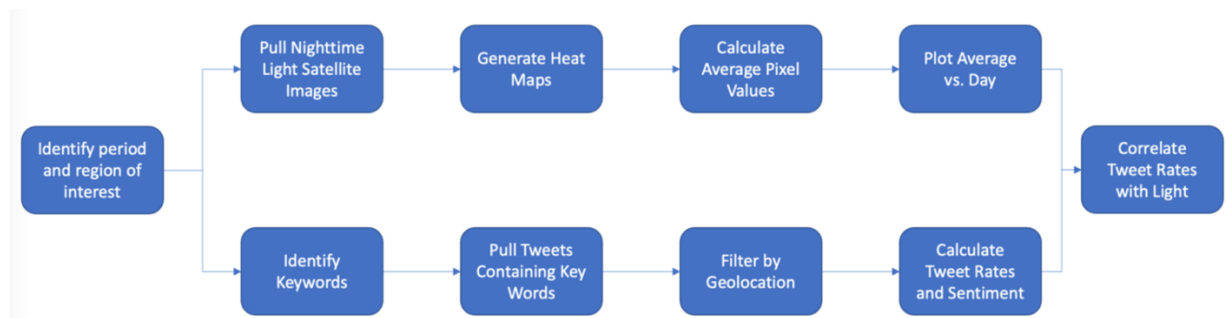

**Figure 3. Flow chart of basic methodology for nighttime satellite image processing and tweet processing and correlation**.

*Nighttime Lights.* Nightly images were obtained for the month of March 2019 from the VIIRS Day/Night Band images distributed by NOAA. Each night contains two orbits, and the orbit with the best coverage of Caracas was selected by determining which of the orbits passed most directly over the region. Once the correct orbit was determined, the region of Caracas was isolated by overlaying a geographically tagged file of Caracas over the satellite image. Pixels within the boundary of the city were kept, while those falling outside the region were set to null. A matrix representation of the image was then created, such that each element represented pixel intensity reported by NOAA in units of nanowatts/cm$^2$/sr. Pixels with a cloud cover value were set to zero, using the VFLAG pixel quality indicator.

A heatmap of the pixel intensity was created. Figure 4 shows the dramatic differences between March 10th and March 12th for the region with these heatmaps. An average pixel value was determined for each day by calculating the mean of all pixel elements with a value greater than 0.05 nW/cm$^2$/sr, a threshold set based on the noise floor of the satellite detectors[34]. Below this value, pixels were set to black (e.g. Fig. 4 – left). Heatmap images covering March 7 through March 13, 2019 are given in the Supplemental Information Note S3, Heatmaps, Caracas.

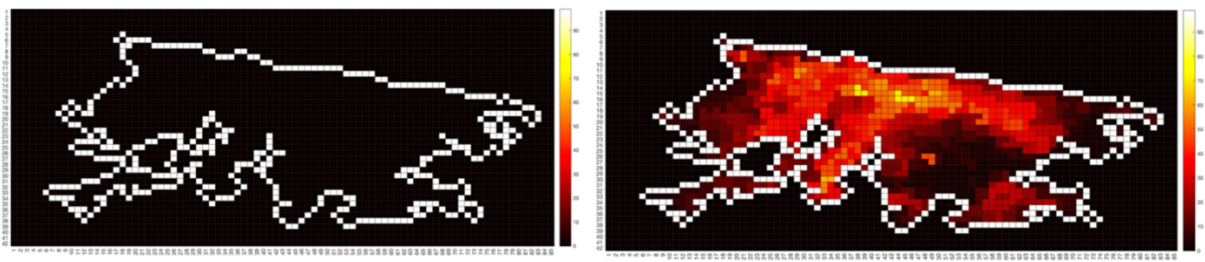

**Figure 4. Caracas, Venezuela: Heatmap of Nighttime Light, March 10$^{th}$ and March 12$^{th}$.** (left) Nighttime light image of Caracas on March 10$^{th}$, during a city-wide blackout. (right) Nighttime light image of Caracas on March 12$^{th}$, when electricity had been restored to the city. Note: The white pixels outline the region of interest and should not be interpreted as light. Here white is the region boundary and lighter colors interior to the boundary indicate more light.

*Social Media Data.* Facebook, Twitter, and Instagram were manually searched for information regarding the blackouts in Caracas and across Venezuela to establish corresponding keywords. The hashtag #sinluz was found to be commonly used to discuss this specific string of outages across these platforms and was used with the Twitter API to identify relevant tweets under an academic license. These tweets were isolated first to Venezuela, then further isolated to those emanating from Caracas and this was done using the user supplied location data. A tweet rate in units of Tweets/day was calculated using the timestamp metadata with each individual tweet.

The tweet dataset was analyzed in the original Spanish to compute the word frequency after discarding stop words (a list of stop words removed from the Spanish tweets is given in Supplementary Note 1). The results, translated to English via google translate, are shown in Table S3 of the Supplementary Information document. Word clouds were generated in English, with the size of the word correlated to the number of counts. From these data, the subset of tweets containing "Maduro" were isolated, and a word frequency for these tweets was generated. The results are shown in Table S4 of the Supplementary Information Note 4. Tweets were filtered into three categories: all tweets, those containing the Venezuelan president's name during this period, "Maduro" (non-case sensitive), and those that did not mention "Maduro". The frequency of "blame," and synonyms for it, were tallied for tweets with and without "Maduro." Here, instances of "Blame," "Irresponsible," "Responsible," "Responsibility," "Irresponsibility," accountable," fault," guilty," and "condemning" were tallied for the tweets geolocated to Venezuela.

Tweets were analyzed in their original Spanish for sentiment by using the Sentiment Spanish package[46]. This is a Python library that uses Keras and Tensorflow to implement a convolutional neural network to determine the sentiment of Spanish sentences. The model comes pretrained on more than 800,000 user reviews from platforms like eltenedor, decathlon, tripadvisor, filmaffinity, and eBay[48]. The Sentiment Spanish package has a reported validation accuracy (on non-training data) of 88%. Each tweet was assigned a score, ranging from 0 to 1, with 0 being the most negative possible, and 1 most positive. Histograms were created to bin the tweets based on sentiment scores in 0.1 increments and using Sturges method within the 0.1 bin. A Spearman's Rank correlation of nighttime lights and tweet rates for Caracas was performed to determine the correlation between power failures and social media response. The non-parametric Mann-Whitney U and Wilcox Signed Rank tests were used to establish the statistical difference in sentiment for tweets containing 'Maduro' and not containing 'Maduro.' The tweets in the lowest sentiment bin were used for this as the number of tweets in other bins was too low. Additional details on the statistical analysis can be found in Supplementary Note S5 and S6.

*Statistical Analysis.* To determine correlation between nighttime light and tweet rates, a Spearman's Rank correlation was used. Non-parameteric Mann-Whitney U and Wilcox Signed Rank tests were used to determine statistical difference in sentiment scores for tweets. Normal statistics of mean, median, mode, standard deviation, max and min were run on tweet scores, and can be found in Supplemental Information, Supplementary Note 2, Statistics. Bin sizes for histograms were determined using Sturges Method, found in Supplemental Information, Supplementary Note 6, Sturges Method.

**Key Resources Table**

| Software or Algorithm | Source | Link |
|---|---|---|
| RAVEN | Idaho National Laboratory | https://raven.inl.gov/SitePages/Overview.aspx |
| VIIRS Nighttime Light Database, Daily | Earth Observation Group, Payne Institute | https://eogdata.mines.edu/nighttime_light/nightly/rade9d/ |
| Sentiment-Spanish | Hugo J. Bello, Universidad de Valladolid | https://github.com/sentiment-analysis-spanish/sentiment-spanish |
| Academic Twitter API | Twitter, Inc. | https://developer.twitter.com/en/use-cases/do-research/academic-research |

**Acknowledgements.** We thank NASA and ESA for the availability of satellite data.

**Author Contributions.** SW Kerber (analyzed data, statistical analysis, wrote and revised the manuscript); Nicholas A Duncan (developed algorithms for sentiment analysis of Twitter data), Guillaume F L'Her (developed algorithms for information extraction and sentiment analysis of Twitter data),, Morgan Bazilian (wrote paper), Chris Elvidge (analysis of nighttime light data), Mark R Deinert (conceived study, wrote and revised paper).

# Supplementary Information
# Tracking electricity losses and their perceived causes using nighttime light and social media

SW Kerber, NA Duncan[2], GF L'Her[1,2], M Bazilian[3], C Elvidge[3], MR Deinert[1,2,3]
Department of Mechanical Engineering[1]
Nuclear Science and Engineering[2]
Payne Institute for Public Policy[3]
The Colorado School of Mines

**Note 1. Excluded word cloud words**

Word clouds excluded normal stop words, such as a/an, the, of, etc. In addition to these words, some other words were removed and are relayed here.

*All tweets not containing "Maduro".*

Words were removed for clarity when creating the word cloud for all tweets. These words were:
https
sinluz
withoutlight
rt

Here, https refers to a hyperlink embedded in the tweet, sinluz is included in all tweets, withoutlight is the simple translation of sinluz and sometimes appears via the translation method but other times does not, rt stands for "retweet" and is excluded because the contents of the retweet are counted rather than the flag.

*All Maduro tweets.*

maduro
sinluz
https
rt
withoutlight
nicolasmaduro
nicol

Maduro is excluded, as well as variations on his name, as that is how these tweets were filtered so all tweets will contain that word. The same words were then removed as in the all tweets word cloud.

**Supplementary Note 2. Statistics**
This section presents various statistics for the tweets presented in the document.

| Bin | Standard Deviation for each bin | | | | | |
|---|---|---|---|---|---|---|
| | All Venezuela Tweets | All Caracas Tweets | Venezuela Maduro Tweets | Caracas Maduro Tweets | Venezuela Excluding Maduro | Caracas excluding Maduro |
| 0.1 | 0.018 | 0.013 | 0.021 | 0.018 | 0.019 | 0.021 |
| 0.2 | 0.029 | 0.031 | 0.028 | 0.053 | 0.029 | 0.027 |
| 0.3 | 0.026 | 0.032 | 0.025 | | 0.025 | 0.025 |
| 0.4 | 0.032 | 0.030 | 0.025 | | 0.032 | 0.027 |
| 0.5 | 0.035 | 0.023 | 0.036 | 0.031 | 0.035 | 0.036 |
| 0.6 | 0.028 | 0.019 | 0.032 | | 0.028 | 0.033 |
| 0.7 | 0.031 | | 0.031 | | 0.031 | 0.031 |
| 0.8 | 0.025 | | 0.029 | | 0.025 | 0.029 |
| 0.9 | 0.031 | 0.008 | 0.027 | 0 | 0.034 | 0.032 |
| More | 0.035 | 0.011 | 0.038 | | 0.035 | 0.038 |

*Table S1. Standard Deviations for sentiment scores falling into each bin.* The categories in this figure correspond to those shown in Table S2.

| Statistics for bin 0.1 | | | | | |
|---|---|---|---|---|---|
| | Mean | Mode | Median | Max | Min |
| *Venezuela* | 8.5E-3 | 1.11E-4 | 2.13E-4 | 9.8E-2 | 5.66E-15 |
| *Venezuela Maduro* | 4.6E-3 | 2.0E-6 | 1.5E-5 | 8.7E-2 | 7.11E-12 |
| *Venezuela Excluding Maduro* | 9.0E-3 | 1.11E-4 | 2.93E-4 | 9.8E-2 | 5.66E-15 |
| *Caracas* | 1.07E-2 | 1.11E-4 | 4.38E-4 | 9.8E-2 | 5.66E-15 |
| *Caracas Maduro* | 9.7E-3 | 4.1E-3 | 5.5E-5 | 7.6E-2 | 7.11E-12 |
| *Caracas Excluding Maduro* | 1.09E-2 | 1.11E-4 | 4.89E-4 | 9.8E-2 | 5.66E-15 |

*Table S2. Various measures for tweets contained in bin 0.1 are presented for the investigated categories.* Note tweets excluding Maduro from Venezuela are two orders of magnitude more positive than those including Maduro in both mean and median.

The word clouds were generated using the browser version of Word Art.

**Note 3. Heatmaps, Caracas**

Heatmaps were created for the Caracas region for March 7 through March 13. These heatmaps

are presented here.

Figures S1 through S7 show these heatmaps.

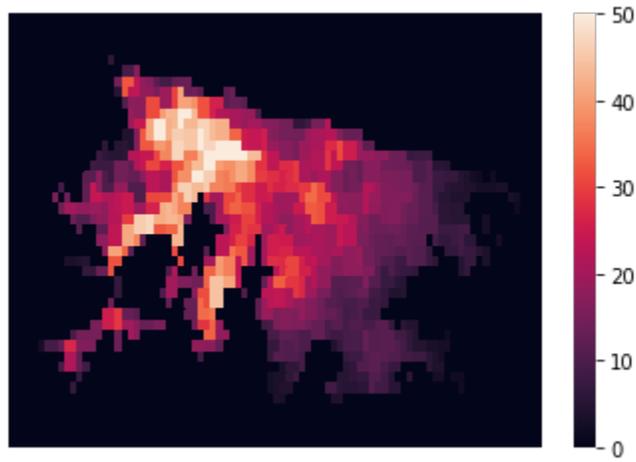

*Figure S1. Caracas Nightlights on March 7*

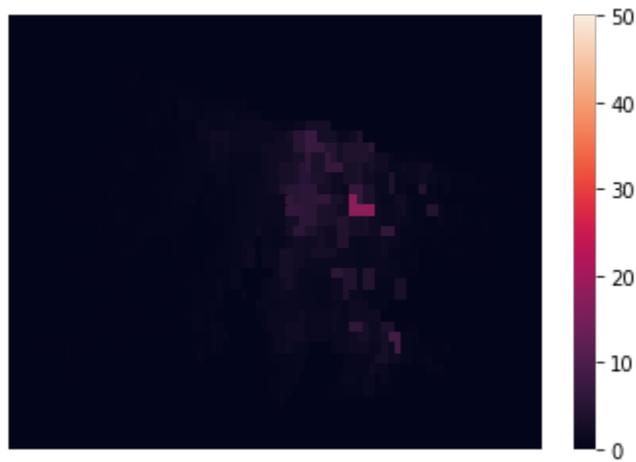

*Figure S2. Caracas Nightlights on March 8*

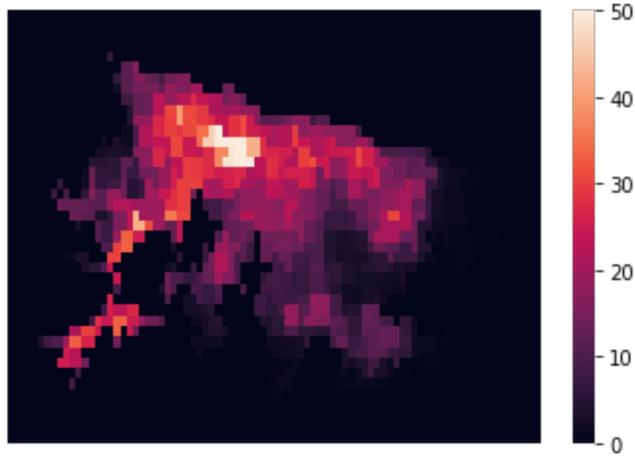

*Figure S3. Caracas Nightlights on March 9*

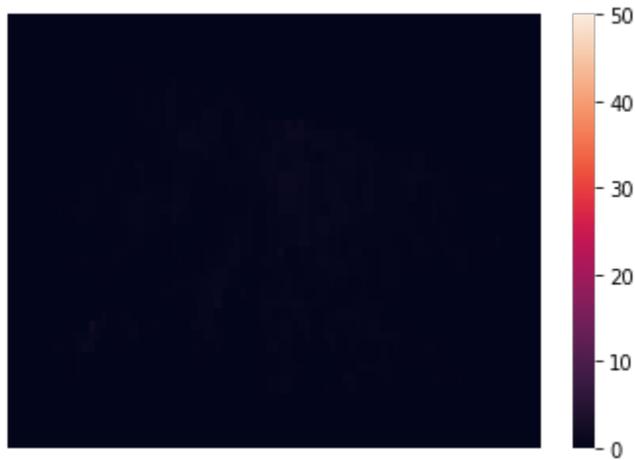

*Figure S4. Caracas Nightlights on March 10*

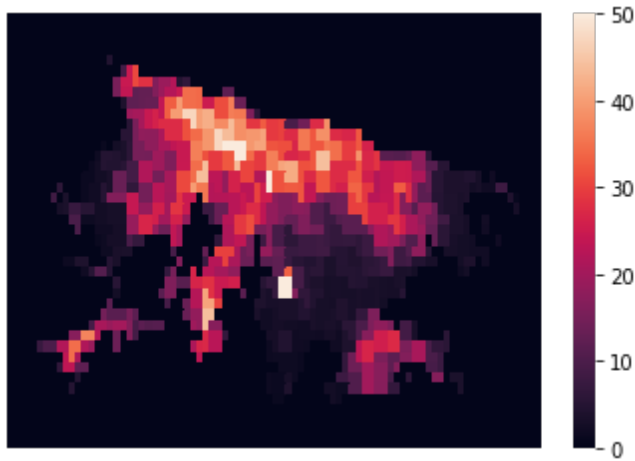

*Figure S5. Caracas Nightlights on March 11*

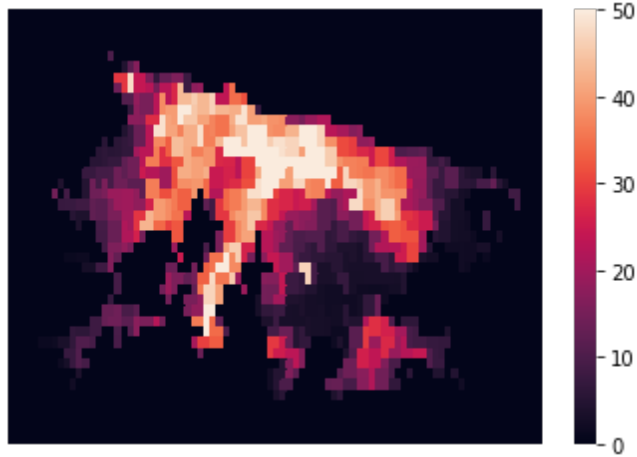

*Figure S6. Caracas Nightlights on March 12*

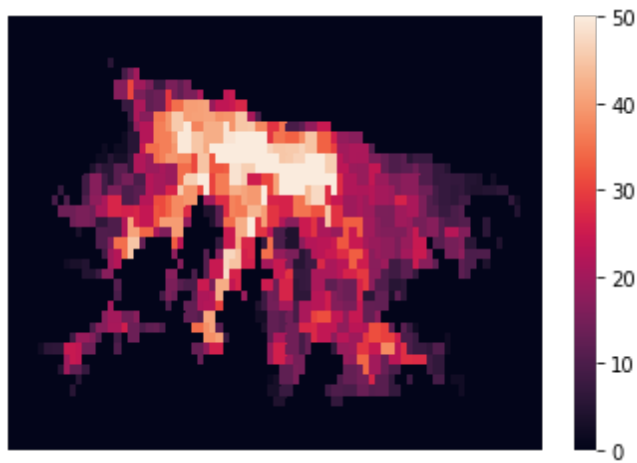

*Figure S7. Caracas Nightlights on March 13*

**Supplementary Note 4. Most used words within tweets.**
This note presents the most frequently used words within all tweets collected and from tweets containing "Maduro." Words were converted to lowercase to get around any case sensitivity issues. Stop words were removed from the Spanish tweets (rt, de, en, la, que, a, y, el, un, se, the, los, una, del, no, https, t, es, las, le, por, debe, con, para, in, lo, and, hay, este, su, on, porque, was, ya, esto, mi, al, of). The results were converted into English using Google Translate and the word frequency computed.

| *Word* | *Total Number of Occurrences* | *% of All Words* |
|---|---:|---:|
| maduro | 131 | 0.25 |
| plant | 118 | 0.22 |
| water | 101 | 0.19 |
| looted | 96 | 0.18 |
| bank | 90 | 0.17 |
| food | 84 | 0.16 |
| megaapagon | 79 | 0.15 |
| sabotage | 79 | 0.15 |
| guard | 69 | 0.13 |
| chaos | 61 | 0.12 |
| repression | 58 | 0.12 |
| supermarkets | 58 | 0.12 |
| energy | 53 | 0.10 |
| died | 53 | 0.10 |
| born | 53 | 0.10 |
| lost | 51 | 0.098 |
| dialysis | 48 | 0.092 |
| deceased | 46 | 0.088 |
| power | 46 | 0.088 |
| potable | 45 | 0.086 |
| hospital | 43 | 0.0824 |

*Table S3. Table of most used words in all tweets.*

| *Word* | *Number of Occurrences* | *% of All Words* |
|---|---|---|
| maduro | 133 | 3.87 |
| maduradascom | 51 | 1.48 |
| hours | 31 | 0.90 |
| mega | 22 | 0.64 |
| caracas | 19 | 0.55 |
| electric | 19 | 0.55 |
| humans | 18 | 0.52 |
| materials | 18 | 0.52 |
| lost | 18 | 0.52 |
| responsible | 18 | 0.52 |
| all | 18 | 0.52 |
| venezuela | 18 | 0.52 |
| nicolasmaduro | 16 | 0.47 |
| hospitals | 16 | 0.47 |
| unforgivable | 15 | 0.44 |
| Rally | 15 | 0.44 |
| died | 15 | 0.44 |

| | | | |
|---|---|---|---|
| kids | | 15 | 0.44 |
| screens | | 15 | 0.44 |
| plant | | 15 | 0.44 |
| sound | | 15 | 0.44 |

*Table S4. Table of most used words found in tweets containing "Maduro."*

**Supplementary Note 5. Histograms of Tweet Sentiment Scores**
This note utilizes histograms to illustrate the distribution of tweet sentiment scores across buckets ranging from 0 to 1.

| | *Tweet histograms as a %* | | | | | |
|---|---|---|---|---|---|---|
| Bin | *All Venezuela Tweets* | *All Caracas Tweets* | *Venezuela Maduro Tweets* | *Caracas Maduro Tweets* | *Venezuela Excluding Maduro* | *Caracas Excluding Maduro* |
| 0.1 | 71.4 | 72.3 | 76.9 | 75.4 | 70.7 | 72.0 |
| 0.2 | 5.2 | 5.7 | 2.9 | 2.9 | 5.5 | 6.0 |
| 0.3 | 4.9 | 4.5 | 5.0 | 1.4 | 4.9 | 4.8 |
| 0.4 | 1.8 | 1.6 | 2.1 | 1.4 | 1.7 | 1.6 |
| 0.5 | 6.3 | 5.9 | 4.1 | 7.2 | 6.6 | 5.7 |
| 0.6 | 1.1 | 1.5 | 1.7 | 1.4 | 1.1 | 1.5 |
| 0.7 | 1.3 | 1.1 | 0.0 | 0.0 | 1.5 | 1.2 |
| 0.8 | 1.8 | 1.6 | 0.4 | 0.0 | 1.9 | 1.8 |
| 0.9 | 2.6 | 2.8 | 4.5 | 8.7 | 2.4 | 2.2 |
| More | 3.7 | 3.1 | 2.5 | 1.4 | 3.8 | 3.2 |

*Table S5. Histograms for Tweet categories showing the percentage of sentiments of tweets falling into each bin. "All Venezuela" and "All Caracas" show sentiment scores for every tweet in those regions. Venezuela/Caracas "Maduro Tweets" shows scores only for tweets containing "Maduro." Venezuela/Caracas "Excluding Maduro" shows scores only for tweets not containing "Maduro."*

| | Histogram of Tweet Counts | | | | | |
|---|---|---|---|---|---|---|
| Bin | *All Venezuela Tweets* | *All Caracas Tweets* | *Venezuela Maduro Tweets* | *Caracas Maduro Tweets* | *Venezuela Excluding Maduro* | *Caracas Excluding Maduro* |
| 0.1 | 1660 | 544 | 186 | 52 | 1474 | 492 |
| 0.2 | 121 | 43 | 7 | 2 | 114 | 41 |
| 0.3 | 114 | 34 | 12 | 1 | 102 | 33 |
| 0.4 | 41 | 12 | 5 | 1 | 36 | 11 |
| 0.5 | 147 | 44 | 10 | 5 | 137 | 39 |
| 0.6 | 26 | 11 | 4 | 1 | 22 | 10 |
| 0.7 | 31 | 8 | 0 | 0 | 31 | 8 |
| 0.8 | 41 | 12 | 1 | 0 | 40 | 12 |
| 0.9 | 60 | 21 | 11 | 6 | 49 | 15 |
| More | 85 | 23 | 6 | 1 | 79 | 22 |

*Table S6. Histogram for Tweet categories showing the number tweets with sentiment scores falling into each bin.*

**Note 6. Sturges Method**

Histograms were created investigating tweets in bin 0.1 for tweets containing Maduro and tweets excluding Maduro using binning determined by Sturges Method. The following equation can be used to determine the number of classes according to this method.

$$K = 3.22 \log N \qquad \text{(Eq.S.1.)}$$

Where K is the number of classes and N is the number of observations.

Using eq.S.1, the width of histogram bins may be calculated using eq.S.2.

$$Width = \frac{Max - Min}{K} \qquad \text{(Eq.S.2)}$$

Where max is the max value (0.1) and min is the minimum value (0.0).

Table S7 shows the number of observations (N), the number of classes (K), and the width for each category.

| | *Venezuela Maduro Tweets Sample Size* | *Caracas Maduro Tweets Sample Size* | *Venezuela Excluding Maduro Sample Size* | *Caracas Excluding Maduro Sample Size* |
|---|---|---|---|---|
| *Bin 0.1* | 186 | 52 | 1474 | 492 |
| *Sturgis K Number of bins* | 9 | 7 | 12 | 10 |
| *Bin Width* | 0.0111111 | 0.0142857 | 0.0083333 | 0.0100000 |

*Table S7. Number of tweets in 0.1 bin by category.*

Tables S8-S11 present tweets falling into bin 0.1 according to Sturges Method for tweets from Venezuela and Caracas including Maduro and for those excluding Maduro.

| Venezuela Excluding Maduro |
|---|

| Bin | Frequency | Percent % |
|---|---|---|
| 0.0083 | 1110 | 75.305 |
| 0.0167 | 134 | 9.091 |
| 0.0250 | 35 | 2.374 |
| 0.0333 | 39 | 2.646 |
| 0.0417 | 25 | 1.696 |
| 0.0500 | 35 | 2.374 |
| 0.0583 | 34 | 2.307 |
| 0.0667 | 23 | 1.560 |
| 0.0750 | 16 | 1.085 |
| 0.0833 | 7 | 0.475 |
| 0.0917 | 4 | 0.271 |
| More | 12 | 0.814 |

*Table S8. Histogram of tweets with score <0.1 from Venezuela excluding "Maduro."*

| Venezuela Including Maduro | | |
|---|---|---|
| Bin | Frequency | Percent % |
| 0.011 | 165 | 88.710 |
| 0.022 | 9 | 4.839 |
| 0.033 | 1 | 0.538 |
| 0.044 | 6 | 3.226 |
| 0.056 | 1 | 0.538 |
| 0.067 | 2 | 1.075 |
| 0.078 | 1 | 0.538 |
| 0.089 | 1 | 0.538 |
| More | 0 | 0 |

*Table S9. Histogram of tweets with score <0.1 from Venezuela including "Maduro."*

| Caracas Excluding Maduro | | |
|---|---|---|
| Bin | Frequency | Percent % |
| 0.01 | 366 | 74.390 |
| 0.02 | 42 | 8.537 |
| 0.03 | 15 | 3.049 |
| 0.04 | 9 | 1.829 |
| 0.05 | 22 | 4.472 |
| 0.06 | 9 | 1.829 |
| 0.07 | 11 | 2.236 |
| 0.08 | 9 | 1.829 |
| 0.09 | 3 | 0.610 |
| More | 6 | 1.220 |

*Table S10. Histogram of tweets with score <0.1 from Caracas excluding "Maduro."*

| Caracas Including Maduro | | |
|---|---|---|
| Bin | *Frequency* | *Percent %* |
| 0.0143 | 40 | 76.923 |
| 0.0286 | 4 | 7.692 |
| 0.0429 | 4 | 7.692 |
| 0.0571 | 2 | 3.846 |
| 0.0714 | 1 | 1.923 |
| 0.0857 | 1 | 1.923 |
| More | 0 | 0 |

*Table S11. Histogram of tweets with score <0.1 from Caracas including "Maduro."*